\documentclass[aps,preprint,showpacs, showkeys]{revtex4}%
\usepackage{amsfonts}
\usepackage{amsmath}
\usepackage{amssymb}
\usepackage{graphicx}
\usepackage{epsfig}
\usepackage{exscale}
\usepackage{float}
\usepackage{bm}
\usepackage{suffix}
\usepackage{mathtools}
\usepackage{newunicodechar}
\usepackage{bbding}
\usepackage{tikz}
\usepackage{multirow}%
\providecommand{\U}[1]{\protect\rule{.1in}{.1in}}
\DeclarePairedDelimiterX\MeijerM[3]{\lparen}{\rparen}
{\begin{smallmatrix}#1 \\ #2\end{smallmatrix}\delimsize\vert\,#3}
\newcommand\MeijerG[8][]{  G^{\,#2,#3}_{#4,#5}\MeijerM[#1]{#6}{#7}{#8}}
\WithSuffix
\newcommand\MeijerG*
[7]{
G^{\,#1,#2}_{#3,#4}\MeijerM*{#5}{#6}{#7}}

\begin{document}
\title[ ]{ The $\lambda$-point anomaly in view of the seven-loop Hypergeometric resummation   for  the
 critical exponent  $\nu$   of the $O(2)$ $\phi^4$ model}
\author{Abouzeid M. Shalaby}
\email{amshalab@qu.edu.qa}
\affiliation{Department of Mathematics, Statistics, and Physics, Qatar University, Al
Tarfa, Doha 2713, Qatar}
\keywords{Critical phenomena, Resummation Algorithms, super-fluid transition }
\pacs{}

\begin{abstract}
In this work, we use a specific parametrization of the hypergeometric approximants ( the one by Mera et.al in Phys. Rev. Let. 115, 143001 (2015)) to approximate the seven-loop critical exponent $\nu$ for the $O(2)$-symmetric $\phi^4$ model.  Our prediction gives the result $\nu=0.6711(7)$ which  is  compatible with  the value $\nu=0.6709(1)$  from the  famous  experiment carried on the space shuttle Columbia. On the other hand, our result is also compatible with recent precise theoretical predictions that are excluding the experimental result. These theoretical results include non-perturbative renormalization group calculations ( $\nu=0.6716(6)$), the   most precise result from Monte Carlo simulations ($\nu=0.67169(7)$)  as well as  the recent conformal bootstrap calculations ($\nu=0.67175(10)$). Although our result is compatible with experiment, the plot of renormalization group  result versus the number of loops suggests that higher orders are expected to add significantly to accuracy and precision  of the $\nu$  exponent in a way that may favor the theoretical predictions. 
\end{abstract}
\maketitle
Slava Rychkov wrote a nice commentary \cite{dispute1} on the recent conformal
bootstrap (CB) prediction for the critical exponent $\nu$ of the O(2) model \cite{dispute} which can describe the $^{4}He$
superfluid phase transition. The CB work in Ref.\cite{dispute} confirmed the most precise result of Monte Carlo (MC) simulations in Ref.\cite{MC19} but excluded the experimental result in Ref.\cite{expermint}. These facts have been summarized  by Slava Rychkov where he outlined the current status of the
predictions of both theoretical and experimental results. The   CB and MC results, in conjunction with   the recent result from  non-perturbative renormalization group (NPRG) \cite{NPRG}, have asserted  what can be called the $\lambda$-point dispute between theory and experiment which lasts for a decade.  Resummation of the perturbation series for the associated renormalization group (RG) function, on the other hand, so far is not precise enough to favor either experiment or the  mentioned non-perturbative  theoretical calculations. 

\begin{figure}[t]
\begin{center}
\epsfig{file=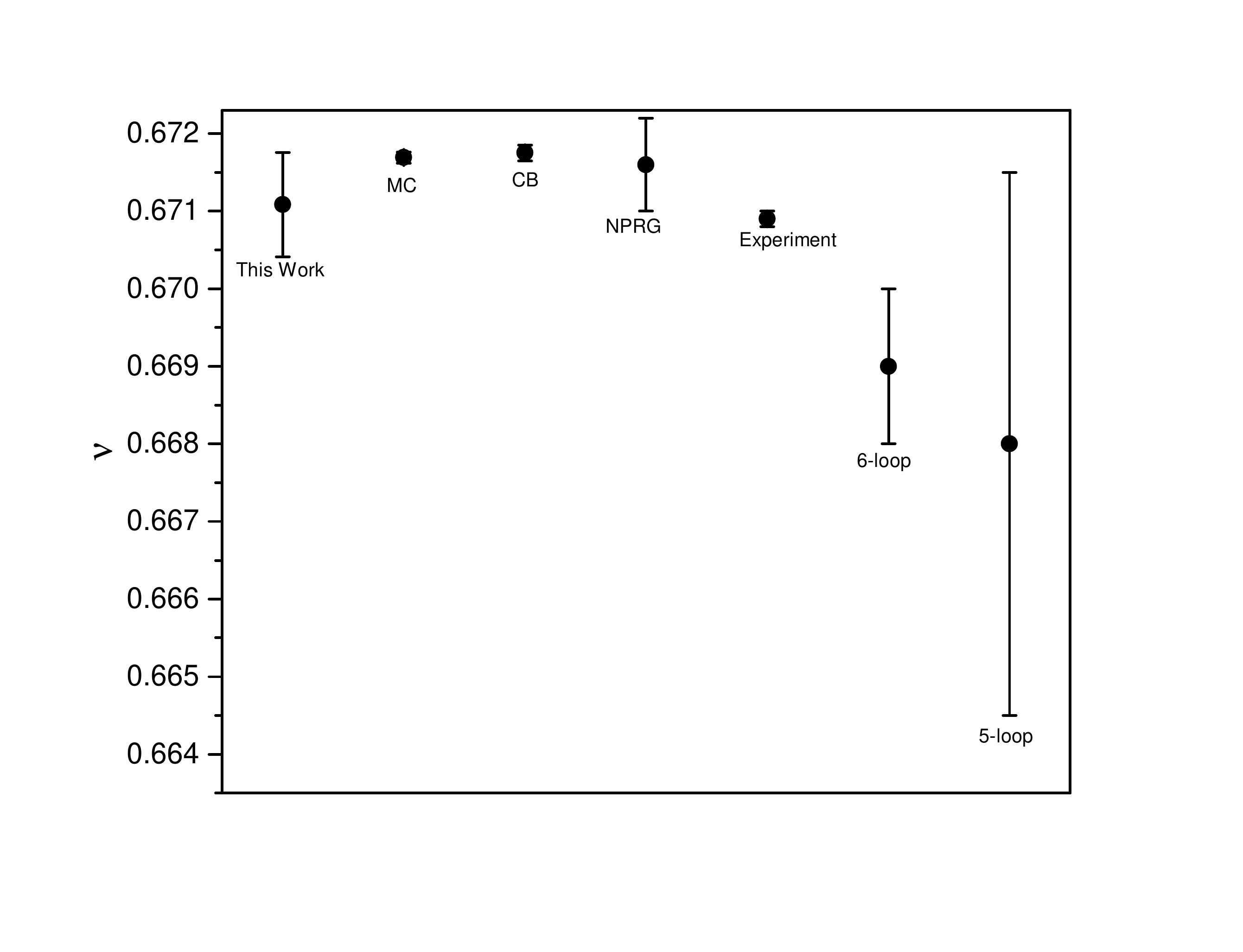,width=0.65\textwidth}
\end{center}
\caption{\textit{In this graph, we show the predictions form seven-loop resummation (this work), the recent Monte Carlo prediction (MC) from Ref.\cite{MC19}, conformal bootstrap result \cite{dispute}, non-perturbative renormalization group (NPRG) \cite{NPRG}, experimental result from Ref.\cite{expermint}, six-loop resummation from Ref.\cite{ON17} and the five-loop result from Ref.\cite{zin-exp}. }} 
\label{Bars}%
\end{figure}

The $\lambda$-shape behavior describing the change in specific heat vs. temperature for  helium  superfluid transition    is characterized by the critical exponent $\alpha$. In fact, the specific heat critical exponent $\alpha$ is related to the exponent $\nu$ by the hyper scaling relation $\alpha=2-3\nu$. Using this relation,  the critical exponent $\nu$ for  the $^{4}He$ super-fluid transition can be extracted from the $\alpha$ value in Ref.\cite{expermint} as $\nu=0.6709(1)$. As mentioned above, this result  is in contradiction with both recent Monte Carlo (MC) simulation result of $\nu=0.67169(7)$ \cite{MC19} and bootstrap calculations of $\nu=0.67175(10)$  \cite{dispute,dispute1}. Note that in Refs.\cite{NPRG,MC20N3}, it has been stated that the recent NPRG result ($\nu=0.6716(6)$) also excludes the experimental result. In other words, while  MC, CB and NPRG predictions  are compatible with each other, they exclude  the result of the famous experiment by Lipa \textit{et. al.} in Ref. \cite{expermint} that leads to the above result for $\nu$ (all these results and others are listed in Fig.\ref{Bars}). 

The best known result from resummation of renormalization group functions at fixed dimension ( $d=3$) is $\nu =0.6703(15)$ \cite{zin-exp}. It is clear that this result is not precise enough and  has not been improved since its appearance in 1998 as explained by Slava Rychkov in Ref.\cite{dispute1}. 
Also, the resummation of the $\varepsilon$-expansion of RG functions is known to have slower convergence than the resummation of RG functions at fixed dimensions \cite{Kleinert-Borel}. This   may explain  the significant difference (see Fig.\ref{Bars}) between theoretical (MC,CB,NPRG) as well as  experimental results  and  the recent six-loop ($\varepsilon$-expansion) resummation result ( $\nu=0.6690(10)$ ) \cite{ON17}. In fact, although  the fixed dimension and the $\varepsilon$-expansion RG results have uncertainties of the same order of magnitude, one can realize that the experimental as well as MC and CB results
exclude the prediction of the six-loop $\varepsilon$-expansion in Ref. \cite{ON17}. In this work, we show that (unlike the six-loop) the seven-loop $\varepsilon$-expansion for the exponent $\nu$ of the $O(2)$-model gives precise result that enables it to play a role in the $\lambda$-point dispute. To do that, we use a modified parametrization of the hypergeometric resummation algorithm \cite{Prl}.

The hypergeometric approximants suggested by Mera \textit{et.al} in Ref. \cite{Prl} uses the
hypergeometric function $_{2}F_{1}(a_{1},a_{2};b_{1};-\sigma x)$ to
approximate a divergent series with $n!$ growth factor (zero radius of
convergence). Although this approximant can give accurate results   \cite{cut,
hyp2, hyp3, hyp4, hyp5, hyp6}, it has been realized (by the same authors) that it suffers from a
genuine problem as the expansion of $_{2}F_{1}(a_{1},a_{2};b_{1};-\sigma x)$
has a finite radius of convergence \cite{cut, hyp3,Prd-GF}. To overcome this issue, the tip of the branch cut is forced to lie at the origin \cite{hyp3,Prd-GF}. In Ref.\cite{abo-hyper}, we have shown that the numerator parameters ($a_{1},a_{2}$) are representing the strong-coupling
asymptotic behavior and thus knowing them can accelerate the convergence. On
the other hand, employing parameters from the large-order asymptotic behavior of the given perturbation series are
well known to accelerate the convergence of resummation algorithms too \cite{Kleinert-Borel,zinjustin}. For the problem under consideration in this work, the large-order parameters are   more important than the strong-coupling parameters as the first are well known in quantum field problems
while the second are not. However, the  approximant $_{2}F_{1}(a_{1}%
,a_{2};b_{1};-\sigma x)$ can not accommodate the large-order parameters because
its expansion does not have the same from of  large-order behavior that the
given perturbation series has. In fact, as stated by Mera \textit{et.al}, the parameter $\sigma$ ought to take large
values to account for the missing $n!$ factorial growth factor \cite{cut,Prd-GF}. Apart from this, one needs to find a way to change the parametrization of the approximant to make it able  to give explicitly the $n!$ growth factor of the given series. 

In Ref.\cite{abohyper-m}, we suggested the parametrization of the form
$_{2}F_{1}(a_{1},a_{2};b_{1};-b_{1}\sigma x)$. If the  parameter  $b_{1}$   takes large
values, then  we have the following limit:
\begin{equation}
\lim_{b_{1}\rightarrow\infty}(_{2}F_{1}(a_{1},a_{2};b_{1};-b_{1}\sigma
x))=\text{ }_{2}F_{0}(a_{1},a_{2};-\sigma x).
\label{limitb1}
\end{equation}
In fact the expansion of $_{2}F_{0}(a_{1},a_{2};-\sigma x)$ has an $n!$ growth
factor \cite{abo-expon} and thus the parameter $\sigma$ can be taken from large-order behavior
of the given series. 

For more explanation of how to apply the new parametrization, consider a quantity $Q$ with the first four
perturbative terms are known:%
\begin{equation}
Q\left(  x\right)  \approx1+\sum_{1}^{3}d_{i}x^{i}.\label{pertQ}%
\end{equation}
This quantity can be approximated by the hypergeometric  approximant $_{2}F_{1}(a_{1},a_{2};b_{1};-b_{1}\sigma x)$ such that: 
\begin{align}
a_{1}a_{2}\sigma &  =d_{1}\nonumber\\
\frac{a_{1}\left(  1+a_{1}\right)  a_{2}\left(  1+a_{2}\right)  }{2\left(
1+b_{1}\right)  }b_{1}\sigma^{2} &  =d_{2}\label{w-algo}\\
\frac{a_{1}\left(  1+a_{1}\right)  \left(  2+a_{1}\right)  a_{2}\left(
1+a_{2}\right)  \left(  2+a_{2}\right)  \left(  b_{1}\right)  ^{2}\sigma^{3}%
}{6\left(  1+b_{1}\right)  \left(  2+b_{1}\right)  } &  =d_{3}.\nonumber
\end{align}
The three unknown parameters ($a_{1},a_{2},b_{1})$ are then obtained by solving the above set of equations.

In case one knows $M+1$ orders, one can approximate the series for
$Q\left(  x\right)  $ by  the generalized hypergeometric function:%
\begin{equation}
Q\left(  x\right)  \approx{}_{k+1}F_{k}(a_{1},a_{2,}........a_{k+1}%
;b,b_{2},....b_{k};-b\sigma x),
\end{equation}
with $M=2k+1$. This strategy has been followed in Ref. \cite{abohyper-m} and
led to accurate results for different examples.

In this work, however, instead of solving   for the parameter $b$, we shall
make it as an input and  set it   as large as possible and then solve for the other
parameters. This setting is very suitable to approach the limit in Eq.(\ref{limitb1}).

Before we use the algorithm to tackle the problem of the critical exponent $\nu$ of the model under consideration, 
let us first apply it to an example for which exact value is known.
To do that, consider the seven-loop of the reciprocal of critical exponent $\nu$
for the Ising case of the $O(N)$-vector model $(N=1)$ \cite{abo-expon}:
\begin{align}
\nu^{-1} &  =2.0000-0.33333\varepsilon-0.11728\varepsilon^{2}%
+0.12453\varepsilon^{3}-0.30685\varepsilon^{4}\label{nueps1}\\
&  +0.95124\varepsilon^{5}-3.5726\varepsilon^{6}+15.287\varepsilon
^{7}.\nonumber
\end{align}
with $\sigma=\frac{1}{3}$ \cite{Kleinert-Borel}. The exact value for
$\varepsilon=2$ (two dimensions ) is $\nu=1$. Taking $b=170$ (the maximum
value one can use due to $\Gamma$ functions defining the coefficients of the
hypergeometric series), one can approximate this series by:
\begin{equation}
\nu^{-1}\left(  \varepsilon\right)  \approx~2.0000-0.33333\ \varepsilon\text{ }%
_{4}F_{3}(a_{1},a_{2},a_{3},a_{4};170,b_{2},b_{3};170\left(  -\frac{1}%
{3}\right)  \varepsilon).
\end{equation}
We find the following values for the parameters: 
\begin{align*}
a_{1}  & =0.0411154,\text{ }a_{2}=13.3176,\text{ }a_{3}=-0.470683,\text{
}a_{4}=-2.29587\\
b_{1}  & =-2.28074,\text{ }b_{2}=0.245782.
\end{align*}
For $\varepsilon=2$, these parameters lead to the result $\nu\left(  2\right)
=0.977203$. To see how our idea improved the  prediction of the algorithm,  we obtained $\nu^{-1}$   by using the original hypergeometric
approximant (with parametrization as presented  in Ref.\cite{Prl}) which has the parametrization:
\begin{equation}
\nu^{-1}\left(  \varepsilon\right)  \approx~2.0000\ -0.33333\ \varepsilon\text{ }%
_{3}F_{2}(a_{1},a_{2},a_{3};b_{1},b_{2};d\text{ }\varepsilon).
\end{equation}
This parametrization gives the result $\nu\left(  2\right)  \approx0.964952$. Note that, the more
sophisticated Borel with conformal mapping resummation for the six-loop of the
same series in Ref.\cite{ON17} gives the result $\nu\left(  2\right)
\approx0.952(14).$  Although the algorithm we follow here in this work might
be the simplest one, it gives  accurate result as shown above. One can
even refine the result by taking larger values for the parameter $b$ but in
this case to overcome the machine limit one can use the limiting case:%

\begin{align*}
\nu^{-1}\left(  \varepsilon\right)    & \approx~\lim_{b\rightarrow\infty}\left(
2.0000-0.33333\varepsilon\text{ }_{4}F_{3}(a_{1},a_{2},a_{3},a_{4}%
;b,b_{2},b_{3};b\left(  -\frac{1}{3}\right)  \varepsilon)\right)  \\
& =\left(  2.0000-0.33333\varepsilon\text{ }_{4}F_{2}(a_{1},a_{2},a_{3}%
,a_{4};\ b_{2},b_{3};\ \left(  -\frac{1}{3}\right)  \varepsilon)\right)  .
\end{align*}
In such a case one has to resort to the representation of the hypergeometric
function $_{4}F_{2}(a_{1},a_{2},a_{3},a_{4};\ b_{2},b_{3};\ \left(  -\frac
{1}{3}\right)  \varepsilon)$ in terms of Meijer-G function \cite{HTF}. \ Using
this and the value of $b=10^{5},$ we get the value $\nu\left(  2\right)
=0.98499.$ This example shows clearly that the simple algorithm we use in this
work can give challenging results and thus is trusted to tackle the  
$\lambda$-point anomaly for $^{4}He$ superfluid phase transition as our main problem.

Now consider the seven-loop critical exponent for the $O(2)$ case  given
by \cite{abo-expon}:
\begin{align}
\nu^{-1} &  =2.0000-0.40000\varepsilon-0.14000\varepsilon^{2}%
+0.12244\varepsilon^{3}-0.30473\varepsilon^{4}\label{nueps2}\\
&  +0.87924\varepsilon^{5}-3.1030\varepsilon^{6}+12.419\varepsilon
^{7},\nonumber
\end{align}
which again can be approximated by ($\sigma=3/10$):
\begin{equation}
\nu^{-1}\left(  \varepsilon\right)  \approx~2.0000-0.40000\varepsilon\text{ }%
_{4}F_{3}(a_{1},a_{2},a_{3},a_{4};b,b_{2},b_{3};b\left(  -\frac{3}{10}\right)
\varepsilon).
\label{7lap}
\end{equation}
again taking $b=10^{5}$ and  $\varepsilon=1$,  we get the result $\nu=0.67094$.
Note that the five-loop resummation for the same series gives $\nu
\ =0.6680(35)$ \cite{zin-exp} while the six-loop resummation gives
$\nu\ =0.6690(10)$ \cite{ON17} and  the experimental result is $0.6709(1)$
\cite{expermint}. Apart from the uncertainty in the calculation which we did not discuss yet, one can realize that our result
shows a significant improvement  of the accuracy of the critical exponent $\nu$ obtained previously from resummation of the $\varepsilon$ expansion  of RG functions. This can be more clarified by comparing with other non-perturbative predictions like the Monte Carlo result which gives  $\nu=0.67169(7)$ \cite{MC19},
  non-perturbative renormalization group (NPRG) that gives  the prediction $\nu=0.6716(6)$ \cite{NPRG}  and the  conformal bootstrap prediction result  $\nu=0.6719(11)$ \cite{Bstrab4}. However, as we mentioned above,  a more precise bootstrap result of $\nu\ =0.67175(10)$ has been recently
appeared in Refs.\cite{dispute1,dispute}.

To know what is the situation of the precision of our resummation result among all other predictions, one needs to offer an estimate for the size of the error in the calculated result. In fact, the sources of errors differ from a theoretical method to  another. For the algorithm we use, the error is due to the unknown higher terms in the perturbation series as well as the arbitrary parameters in the resummation algorithm. Also for MC calculations, errors are due to  Monte Carlo statistical errors and systematic errors associated with the correction to scaling \cite{dispute1}. For other  methods like NPRG and CB they also have their own sources of errors. So it is very natural to have different precision from different methods. An improvement of some calculation can be decided by comparison with previous calculations within the same method. For instance, looking at Fig.\ref{Bars}, one can realize that our seven-loop prediction shows a significant improvement for both accuracy and precession when compared to the five-loop and six-loop predictions.

The algorithm we follow seems to have no arbitrary parameters which one can optimize and thus
find the uncertainty in the result. However, deep understanding of the simple
algorithm we follow can find some implicit arbitrariness in its parametrization. As an example, one can find
different hypergeometric functions that can approximate a given order of
perturbation series. For instance, the seven-loop in Eq.(\ref{nueps2}) can be
approximated by:
\begin{align}
\nu^{-1}\left(  \varepsilon\right)    & \approx~2.0000\text{
}_{5}F_{4}(a_{1},a_{2},a_{3},a_{4},a_{5};b,b_{2},b_{3},b_{4};b\left(  -\sigma\right)
\varepsilon),\nonumber\\
\nu^{-1}\left(  \varepsilon\right)    & \approx~2.0000-0.40000\varepsilon\text{
}_{4}F_{3}(a_{1},a_{2},a_{3},a_{4};b,b_{2},b_{3};b\left(  -\sigma\right)
\varepsilon),\nonumber\\
\nu^{-1}\left(  \varepsilon\right)    & \approx~2.0000-0.40000\varepsilon
-0.14000\varepsilon^{2}+0.12244\varepsilon^{3}\text{ }_{3}F_{2}(a_{1}%
,a_{2},a_{3}\ ;b,b_{2}\ ;b\left(  -\sigma\right)  \varepsilon),\\
\nu^{-1}\left(  \varepsilon\right)    & \approx~2.0000-0.40000\varepsilon
-0.14000\varepsilon^{2}+0.12244\varepsilon^{3}-0.30473\varepsilon^{4}\nonumber\\
& +0.87924\varepsilon^{5}\text{ }_{2}F_{1}(a_{1},a_{2}\ \ ;b\ \ ;b\left(
-\sigma\right)  \varepsilon).\nonumber
\end{align}
 All these approximants are legal and use the same content of
information and having the $n!$ growth factor at the limit $b\rightarrow
\infty$. Of course they give different approximations and the question is which
one shall we select? To answer this question, one also notices that the six-loop
can also be approximated by different hypergeometric functions. A natural  choice
is then to select a pair of approximants for six and seven loops that has the fastest
convergence or equivalently we select the pair that minimizes the difference ($\Delta$) defined as:
\begin{equation}
\Delta=\left\vert \nu_{k}^{7}-\nu_{k^{\prime}}^{6}\right\vert ,
\end{equation}
where $k$ defines the hypergeometric approximant ($_{k}F_{k-1}$) used while superscripts for the number of loops involved. We found
that the approximant:
\[
(\nu^{-1})_{4}^{7}=2.0000-0.40000\varepsilon\text{ }_{4}F_{3}(a_{1},a_{2},a_{3}%
,a_{4};b,b_{2},b_{3};b\left(  -\sigma\right)  \varepsilon),
\]
for the seven-loop and the approximant
\[
(\nu^{-1})_{4}^{6}=2.0000-0.40000\varepsilon-0.14000\varepsilon^{2}\text{ }_{3}%
F_{2}(a_{1},a_{2},a_{3}\ ;b,b_{2}\ ;b\left(  -\sigma\right)  \varepsilon),
\]
for the six-loop give the smallest difference of $\Delta=0.0004.$ This could be used as an
uncertainty and our seven-loop resummation result can be taken as
$\nu=0.67094(4)$. This method of error calculation has been used in
different references ( see for instance sec. 16.6.1 in Ref.\cite{Kleinert-Borel}).

\begin{figure}[t]
\begin{center}
\epsfig{file=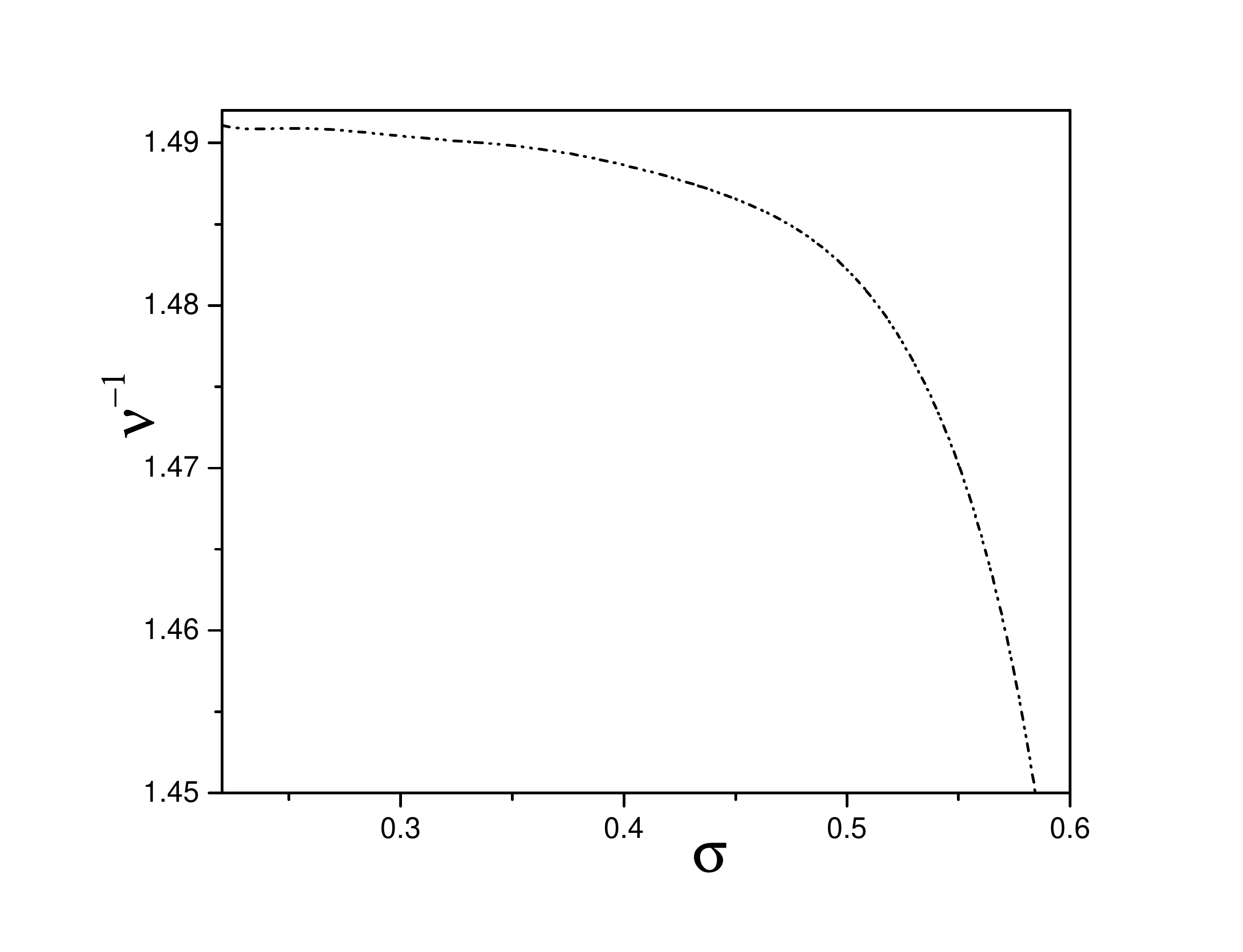,width=0.65\textwidth}
\end{center}
\caption{\textit{The hypergeometric approximant in Eq.(\ref{7lap}) for the seven-loop of $\nu^{-1}$  Vs. $\sigma$ which defines the width $w=x-y$ of the most stable region of the curve. The curve falls down   for $\sigma<0.2$ (not shown in the figure).}} 
\label{nu-vs-sigma}%
\end{figure}

We have another source of arbitrariness  which can be taken from the
form of the large-order behavior of the given perturbation series as:
\begin{align}
Q\left(  x\right)    & \approx\sum_{0}^{n}d_{i}x^{i},\nonumber\\
d_{n}  & =\alpha n!(-\sigma)^{n}n^{b}\left(  1+O\left(  \frac{1}{n}\right)
\right)  ,\text{ \ \ }n\rightarrow\infty.
\end{align}
In fact, the parameter $\alpha$ for the given model is known and has the value
$\alpha=5.892\times10^{-4}$ \cite{large-p3} but none of our equations have been
constrained to account for it. The explicit form of $\alpha$ (Eqs.(4.19,4.10) in Ref.\cite{large-p3}) depends on
$\frac{1}{N+8}=\frac{\sigma}{3}$ which means that one can account for its
variation by varying the parameter $\sigma$ and then find the variance of
$\nu_{4}^{7}$ defined as \cite{ON17}:%
\[
Var_{\sigma}\left(  \nu_{k}^{7}\left(  \sigma\right)  \right)  =\min
_{x\leq\sigma\leq y}\left(  \max_{x\leq\sigma\leq y}\left(\nu_{k}^{7}\left(
\sigma\right)  -\nu_{\acute{k}}^{7}\left(  \sigma^{\prime}\right)  \right)\right).
\]
The width $w=y-x$ is chosen according to the stability region in the curve of 
  the seven-loop exponent $\nu_{k}^{7}\left(  \sigma\right)$ (see Fig.\ref{nu-vs-sigma}). 

Based on the shape of the curve in Fig.\ref{nu-vs-sigma}, we choose $w=0.2$ where we vary $\sigma$ around its exact value of $0.3$ from $0.2$ to $0.4$. 
We use an adapted form of method detailed in Ref.\cite{ON17} where our error can be taken as the minimum of :
\begin{equation}
\Delta\left(  \sigma,k,k^{^{\prime}}\right)  =\left\vert \nu_{k}^{7}\left(
\sigma\right)  -\nu_{k^{\prime}}^{6}\left(  \sigma\right)  \right\vert
+Var_{\sigma}\left(  \nu_{k}^{7}\left(  \sigma\right)  \right)  .
\end{equation}
This form of error is closer in shape to the one used in Ref.\cite{ON17} but adapted in view of the shape in Fig.\ref{nu-vs-sigma} as well as the behavior of six-loop result vs. $\sigma$ (not shown in figures). The error value obtained is  $\Delta_{\min}\approx7\times10^{-4}$ while the predicted value of $\nu$ is $0.6711(7)$. As shown in Fig.\ref{Bars}, this result shows  a significant improvement to the resummation results of the same series in literature. In fact, one can easily realize that moving from six to seven loop, the RG  result is more precise and accurate as it has been shifted toward the MC and BC results.

\begin{figure}[t]
\begin{center}
\epsfig{file=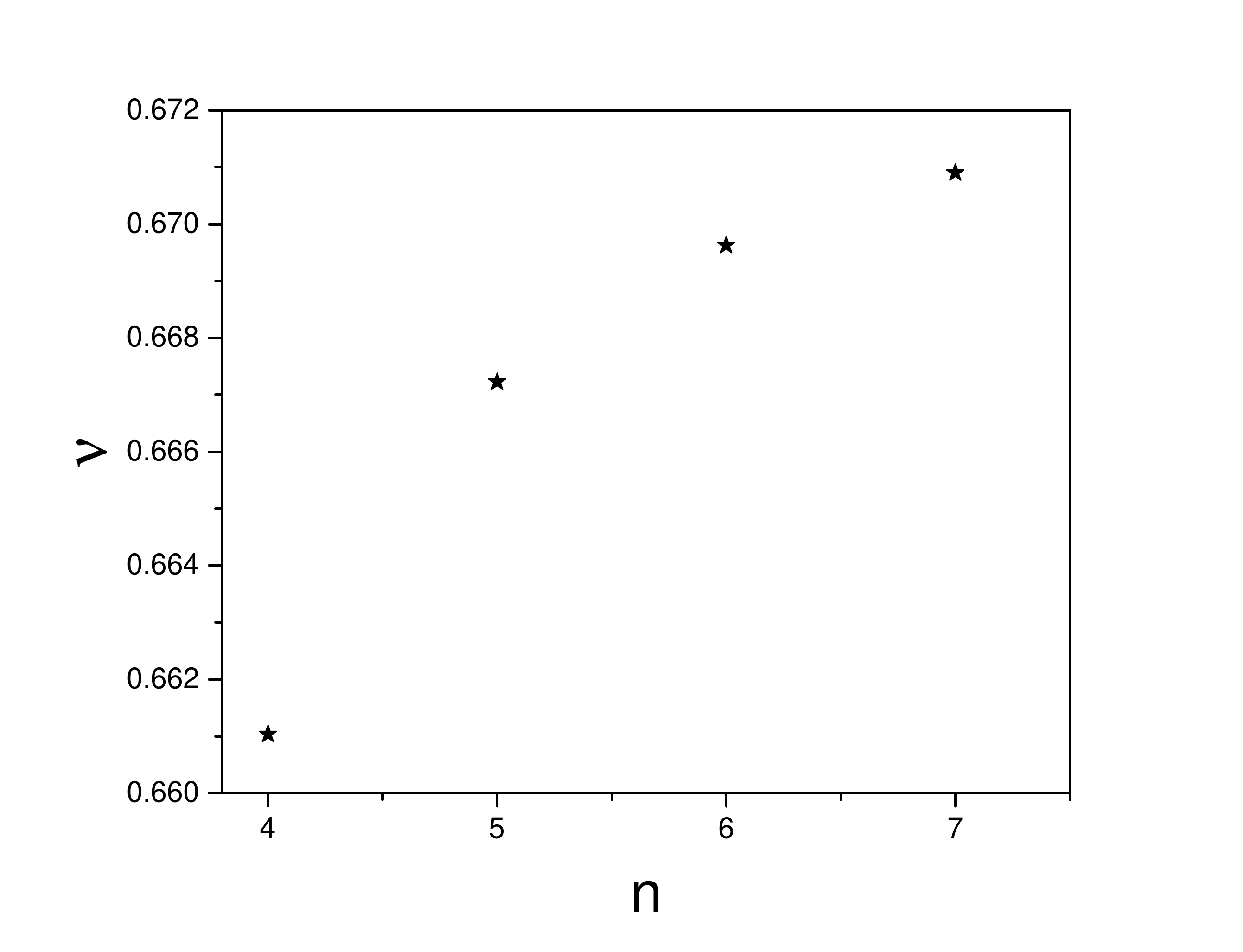,width=0.65\textwidth}
\end{center}
\caption{\textit{In this figure, we plot  our    hypergeometric resummation results for the three dimensional critical exponent $\nu$ of the $O(2)$ model versus the number of loops ($n$) used. One can realize that the distribution of the data   is not saturated enough to claim that the seven-loop result is the best RG result that one can achieve and thus shows the need for more orders.}}
\label{nu-n2}%
\end{figure}

Our prediction is compatible with both experiment as well as MC, BC and NPRG results. However, in view of Fig. \ref{nu-n2}, we see that the seven-loop result is not sufficiently leveling off the curve for $\nu$   versus the number of loops $n$. In
fact, the tangent of the curve is going smaller as function of $n$ but not
small enough to claim a stable shape. This shape of the curve is thus suggesting a possibility 
for smaller error as well as  higher value in the exponent $\nu$ to come from future higher orders. In other words, the
future eight-loop result might exclude the experimental result the same way  NPRG, MC and CB results  do. For a summary of comparison between our results and other methods, we generated table \ref{7L2}.

 \begin{table}[ht]
\caption{{\protect\scriptsize { Our resummation for the seven-loop critical exponent $\nu$ for the $O(2)$ scalar $\phi^4$ model in three dimensions. We list also in the table the famous experimental result from Ref.\cite{expermint}, the most precise result from Monte Carlo  simulations \cite{MC19}, NPRG prediction from Ref.\cite{NPRG} as well as  conformal bootstrap results \cite{dispute,dispute1}. To show the significant improvement that the seven-loop adds to the previous resummation  results of the same series, we list the Borel with conformal mapping (BCM) results for five-loop \cite{zin-exp} and six loops from Ref.\cite{ON17}.\\
} }} 
\label{7L2}
\begin{tabular}{|l|l|l|l|l|l|l|l|}
\hline
Method & $\varepsilon^7$:This work & MC:\cite{MC19} & Experiment:\cite{expermint} & CB: \cite{dispute,dispute1} & NPRG:\cite{NPRG} & $\varepsilon^6$; BCM: \cite{ON17} &$\varepsilon^5$; BCM: \cite{zin-exp} \\ \hline
\ \ \ \ \ $\nu$ & 0.6711(7) & 0.67169(7) & 0.6709(1) & 0.67175(10) &0.6716(6) &0.6690(10) & 0.6680(35) \\ \hline
\end{tabular}
\end{table}

To conclude, we used a simple parametrization of the hypergeometric approximant that enables it to accommodate the large-order parameters for the sake of   convergence acceleration. The modified hypergeometric algorithm is tested first for the two dimensional Ising case where the exact critical exponent is well known. The prediction of the modified hypergeometric algorithm for the seven-loop $\varepsilon$-expansion is very close to the exact result and better than the prediction of the unmodified algorithm as well as the six-loop resummation result from Borel algorithm.

Ironed by the success of the modified hypergeometric algorithm in the two-dimensional case, we tackled the controversial three dimensional case for $\nu$ of the $O(2)$-model. In fact, for $\varepsilon=1$, one expects even better convergence than the two dimensional case. We calculate the exponent $\nu$ for the $O(2)$-symmetric case and get the value $\nu=0.6711(7)$ which is compatible with the experimental result ($\nu=0.6709(1)$) as well as the theoretical calculations from  NPRG method ($\nu=0.6716(6)$), the more precise CB result ($\nu=0.67175(10)$) and MC result  ($\nu=0.67169(7)$). Note that that NPRG, CB and MC results  are excluding the   experimental result.

The plot of the exponent $\nu$ versus the number of loops suggesting that the seven-loop result in this work might not the most precise as well as accurate  prediction that one can obtain from resummation of RG  perturbations. The shape of the curve expecting a more accurate as well as precise result from the future eight-loop series. In other words, there is still a room for RG result to agree with both MC and CB predictions but  excluding the experimental result.
 \section*{Acknowledgment}
We thank Slava Rychkov for drawing our attention to the $\lambda$-point dispute for the $^{4}He$ superfluid phase transition. We also thank  Kay J$\ddot{o}$rg Wiese  for raising to us the possibility of parameter variation for the error estimation.

\end{document}